\documentclass[prb,preprint]{revtex4-1} 

\usepackage{amsmath}
\usepackage{amsfonts}
\usepackage{amssymb}

\usepackage{hyperref}
\usepackage{makeidx}
\usepackage{graphicx}
\usepackage{subcaption}
\usepackage{txfonts}
\usepackage{MnSymbol}
\usepackage{calligra}

\DeclareMathAlphabet{\mathcalligra}{T1}{calligra}{m}{n}
\DeclareFontShape{T1}{calligra}{m}{n}{<->s*[2.2]callig15}{}
\newcommand{\scriptr}{\mathcalligra{r}\,}
\newcommand{\boldscriptr}{\pmb{\mathcalligra{r}}\,}

\begin{document}

\title{Doppler Signature in Electrodynamic Retarded Potentials}

\author{Giovanni Perosa$^{1,2}$, Simone Di Mitri$^{1,2}$, William A Barletta$^{3}$ and Fulvio Parmigiani$^{1,2,4}$}

\email[]{fulvio.parmigiani@elettra.eu}

\address{$^1$ Dipartimento di Fisica, Università di Trieste, via A. Valerio 2, 34127 Trieste, Italy}

\address{$^2$ Elettra Sincrotrone Trieste, I-34149 Basovizza, Trieste, Italy}

\address{$^3$ Dept. of Physics, Massachusetts Institute of Technology Bldg. 24-402, 77 Massachusetts Avenue Cambridge, MA 02139}

\address{{$^4$ International Faculty, University of Cologne, Albertus-Magnus-Platz, 50923 Cologne, Germany}}

\begin{abstract}
\noindent The presence of the term $\left(\case{v}{c}\right)$ that characterizes the electrodynamic retarded potentials, also known as Li\'enard-Wiechert potentials, is thought to be reminiscent of a Doppler effect. Here, we show that these potentials are consistent with the Doppler shifted  electromagnetic (e.m.) field generated by a charge moving along a generic trajectory with respect to the laboratory reference frame. Of course, the retarded potentials derived here are formally the same as those reported in the current literature.  
Nonetheless, this work sheds a new light on the origin of the electrodynamics retarded fields, while offering a direct physical interpretation of the term $\left(\case{v}{c}\right)$ characterizing the related potentials.
\end{abstract}

\maketitle

\section{Introduction}
\begin{figure}[b]
    \centering
    \includegraphics[angle=-90,origin=c,scale=0.35]{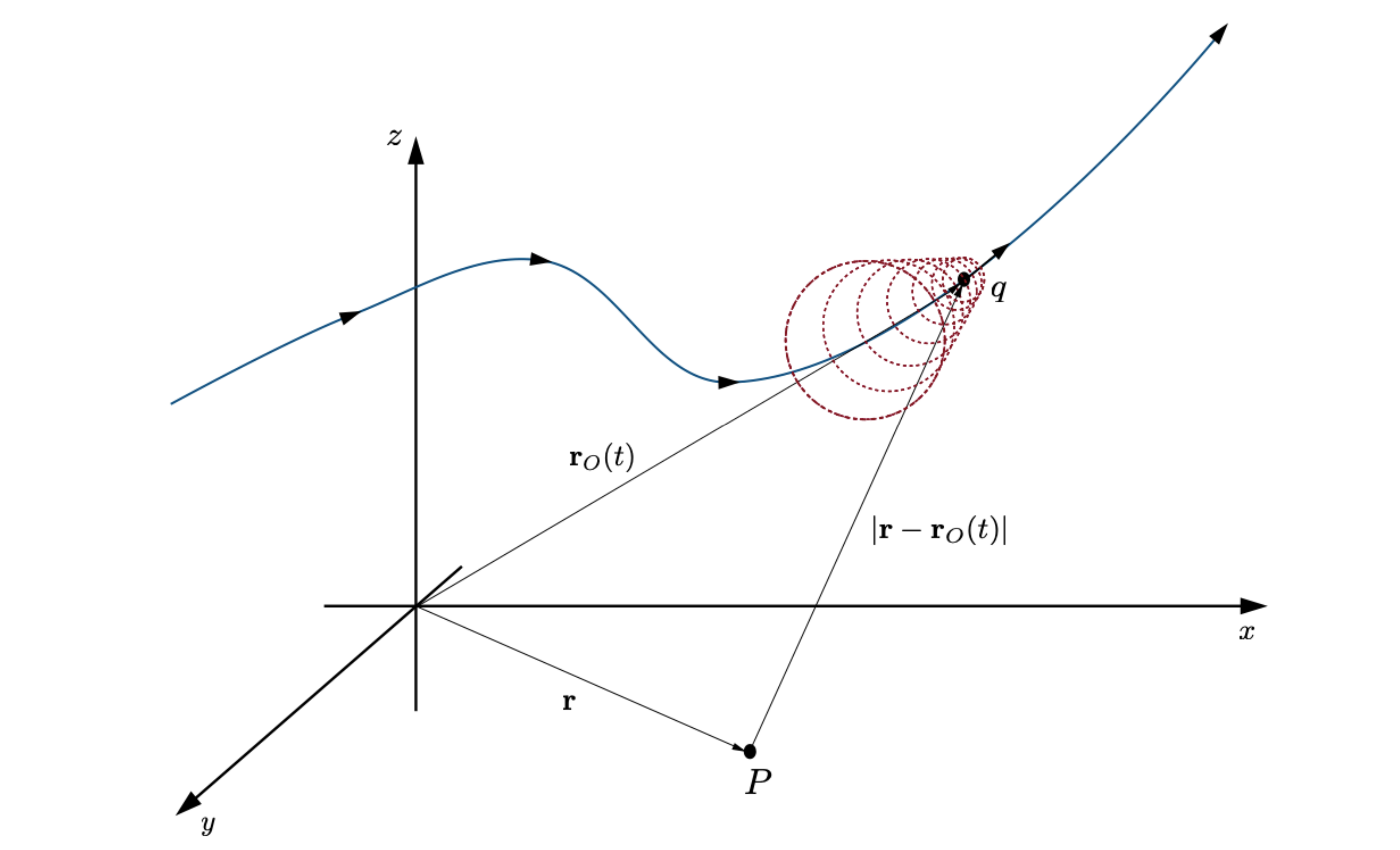}
    \captionsetup{justification=raggedright}
    \caption{Point charge $q$ moving in the reference frame $S$, following the trajectory $\mathbf{r}_O(t)$. The retarded potential at $P$ is computed as the superposition of outgoing spherical wave coming from $q$, schematically represented as circular wavefronts in red. The relative motion of the charge with respect to $P$ is the origin of the Doppler shift of the frequencies.}
    \label{fig1}
\end{figure}

\noindent The retarded potentials, also known as Li\'enard-Wiechert potentials, here after L-W potentials, are the potentials  associated with a charge moving along a generic trajectory, hence of paramount importance in classical and quantum electrodynamics. \cite{Griffith}\\
Assuming that an electromagnetic (e.m.) signal generated by an accelerated charge is traveling at a constant speed, $c$, it will reach a point $\mathbf{r}$, at rest in the laboratory frame, at a later time, $t$, with respect to the starting signal time $t_r$ (see figure. \ref{fig1}). That is why the potentials related to the e.m. field, $\psi(\mathbf{r},t)$ and $\mathbf{A}(\mathbf{r},t)$, are called "retarded potentials". 
Interestingly, these potentials depend on the normalized velocity of the charge, $\left(\case{v}{c}\right) \equiv [\beta\mathbf]$, at the retarded time ($t_r$) and retarded distance respect to the time $t$ and distance $\mathbf{r}$ at which the signal is detected, see point in $P$ in figure. \ref{fig1}.\\

Most textbooks \cite{Jackson,Stratton,Zangwill,Feynman,Panofsky,Orahilly,Reitz,Tommasini} derive the L-W potentials following the original works of Li\'enard  \cite{Lienard} and Wiechert \cite{Wiechert}, which are based on the superposition principle of the fields produced by different volume elements of a charge particle assumed to have a finite size volume. 
Otherwise, L-W potentials are derived via Lorentz transformations, following the hints first given by Minkowski.\cite{Minkowski,Padmanabhan,Streltsov}
Sommerfeld also provided two derivations of the retarded potential by starting from the Green’s function. \cite{Haus,Frank} \\
After Sommerfeld, other authors have derived the L-W potentials using the Green function formalism, see for example the text books of J. Schwinger et al. \cite{Schwinger}, and Zangwill \cite{Zangwill}.\\
The Green's functions formalism is a powerful and elegant mathematical tool, whenever applicable, to describe a physical phenomenon that evolves from an initial state to a final state. However, it leaves the physics hidden, whether it is not properly disclosed.\\
This work has been inspired by a note reported in the widespread textbook of D. J. Griffith where the author, referring to the $\left(\case{v}{c}\right)$ term, writes, \textit{The argument is somewhat reminiscent of the Doppler effect.}\cite{G2}\\
Hence, our aim is to derive the L-W potentials starting from the Doppler effect associated to an e.m. signal generated by an accelerated charge as detected in the laboratory reference frame.\\
In this context, it is remarkable to note that Haus \cite{Haus} derived the e.m. field generated by a single charge moving along a generic trajectory by considering the Fourier-transformed components of the associated current density. He showed that at constant velocity $v<c$, the charge does not radiate because the light-like current density Fourier components are absent. However, while this work helps to elucidate the close relationship between current density and the radiation associated with a moving charge, thus paving the way for an elegant interpretation of the Cherenkov effect, it does not shed light on the Doppler mechanism characterizing the delayed potentials.\\
In this scenario, the difficulty in understanding the nature of the L-W potentials is related to the physical origin of the Doppler-like factor $\left(\case{v}{c}\right)$ that characterises their mathematical representation.  \\
    
\section{Derivation of the retarded potentials}

\subsection{Retarded Green's function}
\noindent Let's start from the retarded potentials in the Lorenz gauge \cite{Griffith, Jackson}
\begin{eqnarray}
     \left( \frac{1}{c^2}\frac{\partial^2}{\partial t^2} - \nabla^2 \right) \psi(\mathbf{r},t) & = & \frac{\rho(\mathbf{r},t)}{\epsilon_0} \label{rho}\\
     \left( \frac{1}{c^2}\frac{\partial^2}{\partial t^2} - \nabla^2 \right) \mathbf{A}(\mathbf{r},t) & = & \mu_0\mathbf{J}(\mathbf{r},t) \label{A}
\end{eqnarray}
where $\psi(\mathbf{r},t)$ is the electric scalar potential, $\mathbf{A}(\mathbf{r},t)$ is the magnetic vector potential, $\rho(\mathbf{r},t)$ is the electric charge density, $\mathbf{J}(\mathbf{r},t)$ is the current density, and $\nabla^2$ is the Laplacian ($\nabla^2 = \frac{\partial ^2}{\partial x^2} + \frac{\partial ^2}{\partial y^2} + \frac{\partial ^2}{\partial z^2}$). These differential equations are often referred to as \textit{wave equations}. \cite{Hassani} The inhomogeneous part, \textit{i.e.} the non trivial right-hand side (RHS), is regarded as the source the electromagnetic waves. In a matter free space, $c$ is the e.m. wave phase velocity. Thereafter, we focus on the case of e.m. fields generated by a moving charge $q$, having charge and current density, 
\begin{equation}
    \rho(\mathbf{r},t) = q\delta(\mathbf{r} - \mathbf{r}_O(t)) \quad \text and\quad {\mathbf{J}(\mathbf{r},t) = q \mathbf{v}(t)\delta(\mathbf{r} - \mathbf{r}_O(t))},
    \label{source}
\end {equation}
respectively, being $\mathbf{r}_O(t)$ the charge trajectory in the laboratory reference frame (as shown in figure. (\ref{fig1})) and $\mathbf{v}(t)$ its velocity.
These equations can be solved by starting from the \textit{retarded Green's function} for the wave equation \cite{Zangwill,Hassani}
\begin{equation}
    G_{-}(\mathbf{r},t) = \frac{1}{4\pi r}\delta(t - r/c).
\end{equation}
Assuming that the observer is affected only by the presence of a source that acts at an earlier time, we can discard the \textit{advanced} solution, proportional to $(t + r/c)$, and use directly the \textit{retarded} one.\\
The Fourier expansion of $G_{-}(\mathbf{r},t)$ in the frequency domain describes a spherical wave propagating outward from the e.m. field source\cite{Zangwill},
\begin{equation}
    G_{-}(\mathbf{r},t) = \frac{1}{2\pi} \int d\omega \Tilde{G}(\mathbf{r},\omega)e^{i\omega t}, \quad \text{with} \quad \Tilde{G}(\mathbf{r},\omega) = \frac{e^{ikr}}{4\pi r}.
    \label{GFF}
\end{equation}

\subsection{Consequences of the relative motion}
\noindent Figure \ref{fig1} shows that, since the signal has a finite velocity ($c = \omega/k$), the information carried by an e.m. wave will take some time to reach the point $P$. 
The time at which the wavefront registered in $P$ at the time $t$ was originated is called retarded time $t_r$. This can be retrieved by the identity of the wavefront phase in $P$ at time $t$ and in the origin at time $t_r$, where the source particle's trajectory is properly taken into account.
Let us indicate with $\phi(\mathbf{r},t)$, the phase of a spherical wave
\begin{equation}
    \phi(\mathbf{r},t) = kr - \omega t,
    \label{phase}
\end{equation}
and the condition for the retarded time translates into
\begin{equation}
    \phi(\mathbf{r},t) = \phi(\mathbf{r}_O (t_r),t_r).
\end{equation}
Inserting the expression for the phase of a spherical phase (\ref{phase}), we obtain:
\begin{equation}
    \phi(\mathbf{r},t) - \phi(\mathbf{r}_O (t_r),t_r) = k|\mathbf{r} - \mathbf{r}_O (t_r)| - \omega(t - t_r) = -\omega\left( t - t_r - \frac{|\mathbf{r} - \mathbf{r}_O(t_r)|}{c} \right) = 0, \nonumber
\end{equation}
hence,
\begin{equation}
    t_r = t - \frac{|\mathbf{r} - \mathbf{r}_O(t_r)|}{c}.
    \label{ret}
\end{equation}
The second relevant consequence of the motion of the source in the reference frame $S$ is the \textit{Doppler shift} of the wave frequency
\begin{eqnarray}
    \omega' = - \frac{d\phi}{dt} & = & - \mathbf{k}\cdot\frac{d \mathbf{r}}{dt} + \omega = \omega \left( 1 - \frac{\hat{\mathbf{k}}\cdot\mathbf{v}}{c} \right) = \frac{\omega}{D_{\hat{\mathbf{k}}}(\mathbf{v})}\\
    \label{doppler}
    D_{\hat{\mathbf{k}}}(\mathbf{v}) & = &  \left( \frac{c}{c - \hat{\mathbf{k}}\cdot\mathbf{v}} \right) \quad \text{and} \quad \hat{\mathbf{k}} = \hat{\mathbf{r}}
\end{eqnarray}

\noindent Equation (\ref{doppler}) can also be interpreted as the change of the phase velocity due to the relative motion of the source in the laboratory reference frame, in a consistent analogy with the apparent pitch change of the sound coming from a moving emitter.\\
A single reference frame has been considered so far, allowing us to avoid any reference to special relativity and relativistic Doppler effect.

\subsection{Computation of the retarded potentials}
\noindent According to the superposition principle, the solution to equation (\ref{rho}) is given by the convolution of the Green's function and the point charge distribution
\begin{equation}
    \psi(\mathbf{r},t) = \frac{1}{\epsilon_0}\int dt' \int d\tau'  G_{-}(\mathbf{r},t|\mathbf{r}',t') \rho(\mathbf{r'},t'),
    \label{psi}
\end{equation}
where $d\tau'$ is the volume three dimensional element and
\begin{equation}
    G_{-}(\mathbf{r},t|\mathbf{r}',t') = \frac{1}{4\pi|\mathbf{r} - \mathbf{r}'|}\delta(t - t' - |\mathbf{r} - \mathbf{r}'|/c).
\end{equation}
In view of the interpretation that we give of the Green's function in the frequency domain, equation (\ref{psi}) represents the superposition of outgoing spherical waves.\\
We can substitute the expression (\ref{GFF}) and the first of equations (\ref{source}) in (\ref{psi}) to get
\begin{equation}
    \psi(\mathbf{r},t) = \frac{q}{2\pi\epsilon_0}\int dt' \int d\tau' \delta(\mathbf{r'} - \mathbf{r}_O(t')) \int d\omega \frac{1}{4\pi|\mathbf{r} - \mathbf{r}'|} \exp{\big\{ i\left[k|\mathbf{r} - \mathbf{r}'| - \omega (t - t')\right] \big\}}.
\end{equation}
The first step is to simplify the inner couple of integrals (the one over $\tau'$ and the one over $\omega$):
\begin{eqnarray}
    \psi(\mathbf{r},t) & = & \frac{q}{2\pi\epsilon_0}\int dt' \int d\tau' \delta(\mathbf{r}' - \mathbf{r}_O(t')) \int d\omega \frac{1}{4\pi|\mathbf{r} - \mathbf{r}'|} \exp{\big\{ i\left[k|\mathbf{r} - \mathbf{r}'| - \omega (t - t')\right] \big\}} = \nonumber \\
    & = & \frac{q}{2\pi\epsilon_0}\int dt' \int d\omega \frac{1}{4\pi|\mathbf{r} - \mathbf{r}_O(t')|} \exp{\big\{ i\left[k|\mathbf{r} - \mathbf{r}_O(t')| - \omega (t - t')\right] \big\}} = \nonumber \\
    & = & \frac{q}{2\pi\epsilon_0}\int dt' \int d\omega \frac{1}{4\pi|\mathbf{r} - \mathbf{r}_O(t')|} \exp{ \big\{i[ \phi(\mathbf{r},t) - \phi(\mathbf{r}_O(t'),t') ] \big\} }
    \label{step1}
\end{eqnarray}
We expand the argument of the exponential, \textit{i.e.} the phase difference, around $t_r$. Actually, the only relevant contributions in the integration over $\omega$ are the ones for which this difference is zero. Clearly, the expansion of the phase difference around $t_r$ is proportional to the phase velocity, which appears to be Doppler shifted, as shown in equation (\ref{doppler}):
\begin{equation}
    \phi(\mathbf{r},t) - \phi(\mathbf{r}_O(t'),t') \approx \frac{\omega}{D_{\hat{\mathbf{k}}}[\mathbf{v}(t_r)]}(t'-t_r) \quad \text{with} \quad \hat{\mathbf{k}} = \frac{\mathbf{r} - \mathbf{r}_O(t_r)}{|\mathbf{r} - \mathbf{r}_O(t_r)|},.
\end{equation}
We will omit the dependence of $\mathbf{v}$ from $t_r$ hereafter.\\
We substitute this new expression in (\ref{step1}),
\begin{eqnarray}
     \psi(\mathbf{r},t) & = & \frac{q}{2\pi\epsilon_0} \int dt'\int d\omega \frac{ e^{i\omega(t' - t_r)/D_{\hat{\mathbf{k}}}(\mathbf{v})} }{4\pi|\mathbf{r} - \mathbf{r}_O(t')|} \nonumber \\
     & = & \frac{q}{2\pi\epsilon_0} \int dt'\int d\omega' \frac{ D_{\hat{\mathbf{k}}}(\mathbf{v}) e^{i\omega'(t' - t_r)} }{4\pi|\mathbf{r} - \mathbf{r}_O(t')|} =  \nonumber \\
     & = & \frac{q}{4\pi\epsilon_0} \int dt' \frac{D_{\hat{\mathbf{k}}}(\mathbf{v})\delta(t' - t_r)}{|\mathbf{r} - \mathbf{r}_O(t')|},
     \label{step3}
\end{eqnarray}
where, the first step is a change of integration variable, from $\omega$ to $\omega'$, while the second step exploits the following identity
\begin{equation}
        \delta(x -x_0) = \frac{1}{2\pi} \int dp e^{ip(x - x_0)}.
\end{equation}
As a result of the integration over $t'$ in equation (\ref{step3}), we find that
    \begin{equation}
        \psi(\mathbf{r},t) = \frac{q}{4\pi\epsilon_0} \frac{ D_{\hat{\mathbf{k}}}(\mathbf{v})}{|\mathbf{r} - \mathbf{r}_O(t_r)|} = \frac{1}{4\pi\epsilon_0}\frac{qc}{(c - \hat{\mathbf{k}}\cdot\mathbf{v}) |\mathbf{r} - \mathbf{r}_O(t_r)|}
    \end{equation}
An analogous strategy can be applied to obtain $\mathbf{A}$, leading to the expression
\begin{equation}
    \mathbf{A}(\mathbf{r},t) = \frac{\mathbf{v}}{c^2}\psi(\mathbf{r},t).
\end{equation}
We can add a few comments to what we have derived so far.
Firstly, it is possible to link our derivation with the one proposed in literature. The evaluation of the charge density at the retarded time is usually expressed as an integration over time of the Dirac delta
    \begin{equation}
    \delta(t' - t_r') = \frac{\delta(t' - t_r)}{|g'(t_r)|} = \frac{1}{2\pi}\int d\nu \frac{e^{i\nu(t' - t_r)}}{|g'(t_r)|} = \frac{1}{2\pi}\int d\omega e^{i\omega(t' - t_r)|g'(t_r)|} \nonumber
    \end{equation}
where we used the fact that
    \begin{equation}
    \delta[g(x)] = \sum_n\frac{\delta(x-x_n)}{|g'(x_n)|},\quad \text{with} \quad g(x_n) = 0, g'(x_n) \neq 0.
    \label{ddf}
    \end{equation}
In this case, $g(t') = t' - t + \scriptr(t')/c$, whose zero is, by definition, the retarded time $t_{r}$. We use the notation $\scriptr(t') = |\mathbf{r} - \mathbf{r}'(t')|$ and find that
\begin{equation}
    g'(t') = \frac{d}{dt'}[t' - t + \scriptr(t')/c] = 1 - \mathbf{v}(t')\cdot\hat{\boldscriptr}/c =  \frac{c - \mathbf{v}(t')\cdot\hat{\boldscriptr} }{c} =  \frac{1}{D_{\hat{\boldscriptr}}(\mathbf{v})}.
\end{equation}
which coincides with our results.\\
Other authors \cite{Zangwill, Panofsky, Orahilly} introduce the Doppler term in different ways and suggest that it can be absorbed in the infinitesimal volume element, stressing with this argument its geometrical origin. In our procedure, instead, that factor is naturally embedded in the treatment and explicitly derived from a Doppler mechanism.\\
More expert scholars could have spotted the presence of a Dirac delta in expression (\ref{step1}) and, using equation (\ref{ddf}), obtaining the final result without passing through all the steps. Although it is undoubtedly correct and more common, we think this haste becomes a lost opportunity to bring out the physics.

\section{Conclusions}
\noindent L-W potentials are a fundamental tool in classical and quantum electrodynamics, although, sometimes the way they are derived hides the physics underneath their origin.\\
Of course, the L-W as reported in the actual literature and textbooks of electrodynamics are formally correct and in many cases exploring the physical mechanism remains an unneeded subtext for their practical use.\\
Nevertheless, there could be cases where the physics underlying L-W potentials could come in handy for the interpretation of processes closely related to L-W potentials.\\
Here we have shown that the L-W potentials can be derived on the basis of a Doppler effect, as suggested by the characteristic correction factor $\left(\case{v}{c}\right)$ presents in their analytical formulation.\\
Overall, the route followed here, besides being formally correct, makes evident the physical phenomenology from which the electrodynamics L-W potentials originate.

\section*{Authors' Declaration}
The authors have no conflicts to disclose.

\appendix

\section{Retarded fields}
\noindent In this appendix we apply the same procedure presented above to find the equations for the retarded fields, or \textit{Jefimenko-Feynman equations}.\cite{Feynman}\\
We start from Maxwell's equations in the form of wave equations
\begin{eqnarray}
    \left( \nabla^2 - \frac{1}{c^2}\frac{\partial^2}{\partial t^2} \right) \mathbf{E} & = & \frac{1}{\epsilon_0}\nabla\rho + \mu_0\frac{\partial \mathbf{J}}{\partial t}\\
    \left( \nabla^2 - \frac{1}{c^2}\frac{\partial^2}{\partial t^2} \right) \mathbf{B} & = & - \nabla \times \mu_0\mathbf{J}
\end{eqnarray}
whose solutions are
\begin{eqnarray}
    \mathbf{E} & = & - \int dt' \int d\tau' G_{-}(\mathbf{r},t|\mathbf{r}',t')\left[ \frac{1}{\epsilon_0}\nabla\rho(\mathbf{r}',t') + \mu_0\frac{\partial \mathbf{J}(\mathbf{r}',t') }{\partial t} \right]\\
    \mathbf{B} & = & \int dt' \int d\tau' G_{-}(\mathbf{r},t|\mathbf{r}',t')\left[ \nabla \times \mu_0\mathbf{J}(\mathbf{r}',t') \right].
\end{eqnarray}
Since the operators $\nabla$ and $\partial/\partial t$ act on the unprimed coordinates, for the electric field we have that
\begin{equation}
    \mathbf{E} = - \frac{1}{4\pi\epsilon_0} \nabla \left( \int dt' \int d\tau' G_{-}(\mathbf{r},t|\mathbf{r}',t')\rho(\mathbf{r}',t') \right) - \frac{\mu_0}{4\pi} \frac{\partial}{\partial t} \left( \int dt' \int d\tau' G_{-}(\mathbf{r},t|\mathbf{r}',t')\mathbf{J}(\mathbf{r}',t') \right),
    \nonumber
\end{equation}
while for the magnetic field,
\begin{equation}
    \mathbf{B} = \mu_0 \nabla \times \left( \int dt' \int d\tau' G_{-}(\mathbf{r},t|\mathbf{r}',t')\mathbf{J}(\mathbf{r}',t') \right). \nonumber
\end{equation}
The terms in parenthesis, called $I$ and $\mathbf{\Pi}$ hereafter, are exactly the same integrals already encountered, when the definition (\ref{source}) is inserted. The remaining step is the evaluation of the derivatives with respect to space and time
\begin{equation}
    \mathbf{E} = - \frac{1}{4\pi\epsilon_0} \nabla I(\mathbf{r},t) - \frac{\mu_0}{4\pi} \frac{\partial}{\partial t} \mathbf{\Pi}(\mathbf{r},t),\quad \mathbf{B} = \mu_0 \nabla \times \mathbf{\Pi}(\mathbf{r},t)
    \label{step2}
\end{equation}
Following the steps proposed by Griffith \cite{Griffith}, we need some useful identities to carry out the derivation. Writing equation (\ref{ret}) as $\scriptr(t_r) = c(t-t_r)$, it is easy to show that
\begin{equation}
    \nabla \scriptr = -c\nabla t_r = \nabla(\sqrt{\boldscriptr \cdot \boldscriptr}) = \frac{1}{2\sqrt{\boldscriptr \cdot \boldscriptr}} \nabla (\boldscriptr \cdot \boldscriptr) = \frac{1}{\scriptr}\boldscriptr\cdot(\nabla\boldscriptr),
    \nonumber
\end{equation}
multiplying both sides for $\scriptr$ and considering the $i$-th component
\begin{equation}
    -c\scriptr \frac{\partial t_r}{\partial x_i} = \scriptr_j \frac{\partial \scriptr^j}{\partial x_i} = \scriptr_j\left(\delta_{i}^j - v^j \frac{\partial t_r}{\partial x_i}\right) = \scriptr_i - \scriptr_j v^j\frac{\partial t_r}{\partial x_i}
    \nonumber
\end{equation}
or, in vectorial form,
\begin{equation}
    -c\scriptr\nabla t_r = \boldscriptr - (\boldscriptr\cdot\mathbf{v})\nabla t_r,
    \nonumber
\end{equation}
hence
\begin{equation}
    \nabla \scriptr = - c \nabla t_r = \frac{c}{ c - \hat{\boldscriptr}\cdot\mathbf{v}}\hat{\boldscriptr} = D_{\hat{\boldscriptr}}(\mathbf{v})\hat{\boldscriptr}.
    \label{33}
\end{equation}
Moreover,
\begin{align}
    & c\left( 1 - \frac{\partial t_r}{\partial t} \right) =  \frac{\partial }{\partial t}c(t-t_r) = \frac{\partial \scriptr}{\partial t} = \frac{\partial (\sqrt{\boldscriptr\cdot\boldscriptr})}{\partial t} = \frac{1}{\scriptr}\boldscriptr\cdot\frac{\partial \boldscriptr}{\partial t} = -\frac{1}{\scriptr}\boldscriptr\cdot\mathbf{v}\frac{\partial t_r}{\partial t},
    \nonumber
\end{align}
and rearranging the first and last terms, we get
\begin{equation}
    \scriptr c = (\scriptr c - \boldscriptr\cdot\mathbf{v}) \frac{\partial t_r}{\partial t}, \quad \text{hence} \quad \frac{\partial t_r}{\partial t} = D_{\hat{\boldscriptr}}(\mathbf{v}).
\end{equation}
Given a vectorial quantity $\mathbf{w}(t_r)$, evaluated at the retarded time, the directional derivative along a vector $\mathbf{d}$ and its curl are
\begin{equation}
    (\mathbf{d} \cdot \nabla) \mathbf{w}(t_r) = \frac{d \mathbf{w}(t_r)}{d t_r} \left( \mathbf{d} \cdot \nabla t_r \right), \quad \nabla \times \mathbf{w}(t_r) = - \frac{d \mathbf{w}(t_r)}{d t_r} \times \nabla t_r.
\end{equation}
and using (\ref{33}),
\begin{equation}
    (\mathbf{d} \cdot \nabla) \mathbf{w}(t_r) = -\frac{D_{\hat{\boldscriptr}}(\mathbf{v})}{c}\frac{d \mathbf{w}(t_r)}{d t_r} \left( \mathbf{d} \cdot \hat{\boldscriptr} \right), \quad \nabla \times \mathbf{w}(t_r) = \frac{D_{\hat{\boldscriptr}}(\mathbf{v})}{c} \frac{d \mathbf{w}(t_r)}{d t_r} \times \hat{\boldscriptr}.
    \label{id2}
\end{equation}
We start from the gradient of $I$:
\begin{eqnarray}
    \nabla I(\mathbf{r},t) & = & \nabla \left[ \frac{  D_{\hat{\mathbf{\scriptr}}}(\mathbf{v}(t_r))}{\scriptr} \right] = D_{\hat{\mathbf{\scriptr}}}(\mathbf{v}(t_r)) \nabla \left( \frac{1}{\scriptr} \right) + \frac{\nabla D_{\hat{\mathbf{\scriptr}}}(\mathbf{v}(t_r))}{\scriptr} = \nonumber \\
    & = & -D_{\hat{\mathbf{\scriptr}}}(\mathbf{v}(t_r))\frac{\nabla\scriptr}{\scriptr^2} + \frac{\nabla D_{\hat{\mathbf{\scriptr}}}(\mathbf{v}(t_r))}{\scriptr} = \nonumber \\
    & = & -\frac{D^2_{\hat{\mathbf{\scriptr}}}(\mathbf{v}(t_r))\hat{\boldscriptr}}{ \scriptr^2} + \frac{\nabla D_{\hat{\mathbf{\scriptr}}}(\mathbf{v}(t_r))}{\scriptr} \nonumber 
\end{eqnarray}
We have to simplify the term $\nabla D_{\hat{\mathbf{\scriptr}}}(\mathbf{v})$, which leads to
\begin{equation}
    \nabla D_{\hat{\mathbf{\scriptr}}}(\mathbf{v}) = \nabla\left( \frac{c}{c - \hat{\mathbf{\scriptr}}\cdot\mathbf{v}} \right) = \frac{1}{c}\left( \frac{c}{c - \hat{\mathbf{\scriptr}}\cdot\mathbf{v}} \right)^2\nabla(\hat{\mathbf{\scriptr}}\cdot\mathbf{v}) = \frac{1}{c}D^2_{\hat{\mathbf{\scriptr}}}(\mathbf{v})\nabla(\hat{\mathbf{\scriptr}}\cdot\mathbf{v}).
    \nonumber
\end{equation}
Now,
\begin{eqnarray}
    \nabla(\hat{\mathbf{\scriptr}}\cdot\mathbf{v}) & = & \nabla \left( \frac{\boldscriptr}{\scriptr}\cdot \mathbf{v} \right) = \nabla \left( \frac{1}{\scriptr} \right) \left( \boldscriptr \cdot \mathbf{v} \right) + \frac{1}{\scriptr} (\nabla(\boldscriptr\cdot\mathbf{v})) = \nonumber \\ 
    & = & \frac{1}{\scriptr} \left[ -\frac{\nabla \scriptr}{\scriptr} \left( \boldscriptr \cdot \mathbf{v} \right) + (\nabla(\boldscriptr\cdot\mathbf{v})) \right] \nonumber\\
    & = &  \frac{1}{\scriptr} \left[ - \frac{D_{\hat{\mathbf{\scriptr}}}(\mathbf{v})\hat{\boldscriptr}}{\scriptr}\left( \boldscriptr \cdot \mathbf{v} \right) + (\nabla(\boldscriptr\cdot\mathbf{v})) \right] \nonumber
\end{eqnarray}
We can exploit the vectorial identities (\ref{id2}) to compute $\nabla(\mathbf{\scriptr}\cdot\mathbf{v})$ and get
\begin{equation}
    \nabla(\mathbf{\scriptr}\cdot\mathbf{v}) = \mathbf{v} -\frac{1}{c} (\boldscriptr\cdot\mathbf{a} -v^2)D_{\hat{\boldscriptr}}(\mathbf{v})\hat{\boldscriptr}.
\end{equation}
Collecting all together the steps, we arrive at
\begin{equation}
    \nabla I(\mathbf{r},t) = -\frac{ D^2_{\hat{\mathbf{\scriptr}}}(\mathbf{v})}{\scriptr^2} \left\{ \hat{\boldscriptr} - \frac{1}{c}\left[ \mathbf{v} -\frac{1}{c} (\boldscriptr\cdot\mathbf{a} -v^2)D_{\hat{\boldscriptr}}(\mathbf{v})\hat{\boldscriptr} - \frac{D_{\hat{\mathbf{\scriptr}}}(\mathbf{v})\hat{\boldscriptr}}{\scriptr}\left( \boldscriptr \cdot \mathbf{v} \right) \right] \right\} \nonumber.
\end{equation}
The first and last terms in the parenthesis simplify into
\begin{equation}
    1 + \frac{\left( \boldscriptr \cdot \mathbf{v} \right)}{c\scriptr}D_{\hat{\mathbf{\scriptr}}}(\mathbf{v}) = \left[ 1 + \frac{(\hat{\boldscriptr}\cdot\mathbf{v})}{c}\frac{c}{c - \hat{\boldscriptr}\cdot\mathbf{v}} \right] = D_{\hat{\mathbf{\scriptr}}}(\mathbf{v}) \nonumber
\end{equation}
and finally,
\begin{eqnarray}
    \nabla I(\mathbf{r},t) & = & -\frac{ D^2_{\hat{\mathbf{\scriptr}}}(\mathbf{v})}{ \scriptr^2}\left\{ D_{\hat{\mathbf{\scriptr}}}(\mathbf{v}) \hat{\boldscriptr} - \frac{1}{c}\left[ \mathbf{v} - \frac{1}{c}(\boldscriptr\cdot\mathbf{a} -v^2)D_{\hat{\boldscriptr}}(\mathbf{v})\hat{\boldscriptr} \right] \right\} = \nonumber \\
    & = & \frac{D^2_{\hat{\mathbf{\scriptr}}}(\mathbf{v})}{\scriptr^2}\left[ \frac{\mathbf{v}}{c} - \frac{D_{\hat{\mathbf{\scriptr}}}(\mathbf{v})}{c^2}(c^2 - v^2 + \boldscriptr\cdot\mathbf{a} )\hat{\boldscriptr} \right].
\end{eqnarray}
Similarly, one can determine $\partial_t\mathbf{\Pi}$,
\begin{eqnarray}
    \frac{\partial \mathbf{\Pi}(\mathbf{r},t)}{\partial t} & = &  \frac{\partial }{\partial t}[\mathbf{v} I(\mathbf{r},t)] = \left[ I(\mathbf{r},t)\frac{\partial t_r}{\partial t}\frac{\partial \mathbf{v}}{\partial t_r} + \mathbf{v}\frac{\partial I(\mathbf{r},t) }{\partial t} \right] = \left[ I(\mathbf{r},t) D_{\hat{\mathbf{\scriptr}}}(\mathbf{v}) \mathbf{a} + \mathbf{v}\frac{\partial I (\mathbf{r},t)}{\partial t} \right] = \nonumber\\
    & = & \left[ \frac{D^2_{\hat{\mathbf{\scriptr}}}(\mathbf{v})}{\scriptr}\mathbf{a} + \mathbf{v}\frac{\partial}{\partial t} \left( \frac{D_{\hat{\mathbf{\scriptr}}}(\mathbf{v})}{\scriptr} \right) \right] = \left[ \frac{D^2_{\hat{\mathbf{\scriptr}}}(\mathbf{v})}{\scriptr}\mathbf{a} + \frac{\mathbf{v}}{\scriptr}\frac{\partial D_{\hat{\mathbf{\scriptr}}}(\mathbf{v})}{\partial t} + \mathbf{v}D_{\hat{\mathbf{\scriptr}}}(\mathbf{v})\frac{\partial}{\partial t}\left( \frac{1}{\scriptr} \right) \right]. \nonumber
\end{eqnarray}
For the second term in RHS,
\begin{eqnarray}
    \frac{\partial D_{\hat{\mathbf{\scriptr}}}(\mathbf{v})}{\partial t} & = & \frac{\partial}{\partial t} \left( \frac{c}{c - \mathbf{v}\cdot\hat{\boldscriptr}} \right) = \frac{1}{c} D^2_{\hat{\mathbf{\scriptr}}}(\mathbf{v}) \frac{\partial }{\partial t}(\mathbf{v}\cdot\hat{\boldscriptr}) = \frac{1}{c} D^2_{\hat{\mathbf{\scriptr}}}(\mathbf{v}) \frac{\partial }{\partial t}\left(\mathbf{v}\cdot\frac{\boldscriptr}{\scriptr}\right) = \nonumber \\  
    & = & \frac{1}{c} D^2_{\hat{\mathbf{\scriptr}}}(\mathbf{v})(\mathbf{v}\cdot\boldscriptr)\frac{\partial}{\partial t}\left( \frac{1}{\scriptr} \right) + \frac{1}{c\scriptr} D^2_{\hat{\mathbf{\scriptr}}}(\mathbf{v}) \frac{\partial }{\partial t}(\mathbf{v}\cdot\boldscriptr), \nonumber
\end{eqnarray}
in which
\begin{align}
    \frac{\partial}{\partial t}\left( \frac{1}{\scriptr} \right) = -\frac{c}{\scriptr^2}[1 - D_{\hat{\mathbf{\scriptr}}}(\mathbf{v})], \quad \text{and} \quad \frac{\partial }{\partial t}(\mathbf{v}\cdot\boldscriptr) = \left( \mathbf{a}\cdot\boldscriptr - v^2 \right) D_{\hat{\mathbf{\scriptr}}}(\mathbf{v}).
\end{align}
The time derivative is
\begin{equation}
    \frac{\partial \mathbf{\Pi}(\mathbf{r},t)}{\partial t} = \frac{D^2_{\hat{\mathbf{\scriptr}}}(\mathbf{v})}{\scriptr^2}\left[ \scriptr \mathbf{a} - c\mathbf{v}  + \frac{\mathbf{v}D_{\hat{\mathbf{\scriptr}}}(\mathbf{v})}{c}\left( c^2- v^2 + \mathbf{a}\cdot\boldscriptr \right)\right]
\end{equation}
As explained in \cite{Griffith}, the calculation of $\nabla \times \mathbf{\Pi}$ can be simplified using the expressions already presented.\\
The final results are
\begin{eqnarray}
    \mathbf{E} & = & -\frac{q}{4\pi\epsilon_0}\left( \nabla I(\mathbf{r},t) +  \frac{1}{c^2}\frac{\partial \mathbf{\Pi}(\mathbf{r},t)}{\partial t} \right) = -\frac{q D^3_{\hat{\mathbf{\scriptr}}}(\mathbf{v})}{4\pi\epsilon_0\scriptr^2}
    \left[ (c^2 - v^2)\mathbf{u} + \boldscriptr\times(\mathbf{u}\times\mathbf{a}) \right]\\
    \mathbf{B} & = & \frac{1}{c}\hat{\boldscriptr}\times\mathbf{E}(\mathbf{r},t),
\end{eqnarray}
where $\mathbf{u} = \hat{\boldscriptr} - \mathbf{v}/c$. Notice that, if we define the new directions
\begin{eqnarray}
    \mathbf{u}_{\parallel} & = & (\mathbf{u}\cdot\hat{\boldscriptr})\hat{\boldscriptr} = D^{-1}_{\hat{\mathbf{\scriptr}}}(\mathbf{v})\hat{\boldscriptr}\\
    \mathbf{u}_{\perp} & = & \mathbf{u} - \mathbf{u}_{\parallel} = \frac{1}{c}\left[ (\hat{\boldscriptr}\cdot\mathbf{v})\hat{\boldscriptr} - \mathbf{v} \right],
\end{eqnarray}
the field along $\mathbf{u}_{\parallel}$ scales as $D^{2}_{\hat{\mathbf{\scriptr}}}(\mathbf{v})/\scriptr^2$, while in the other direction, namely $\mathbf{u}_{\perp}$, is proportional to $D^{3}_{\hat{\mathbf{\scriptr}}}(\mathbf{v})/\scriptr^2$. The latter corresponds to the inverse direction of the velocity component perpendicular to the source position.\\
As suggested in \cite{Heras1}, it is possible to retrieve the retarded potentials from equations (\ref{step2}), invoking the retarded Helmholtz theorem \cite{Heras2}.

\end{document}